\documentclass[preprint]{aastex}

\usepackage{bm}

\DeclareMathAlphabet{\tens}{OT1}{cmss}{bx}{n}

\begin{document}

\title{Removal of Spectro-Polarimetric Fringes by 2D Pattern Recognition}

\author{R.\ Casini,$^1$ P.\ G.\ Judge,$^1$ T.\ A.\ Schad,$^2$}

\affil{%
$^1$ High Altitude Observatory, NCAR,\footnote{The National Center 
for Atmospheric Research (NCAR) is sponsored by the National 
Science Foundation.}
P.\ O.\ Box 3000, Boulder, CO 80307-3000, U.S.A.\newline
$^2$ Lunar and Planetary Lab, University of Arizona,
Tucson, AZ 85721, U.S.A.}

\begin{abstract}
We present a pattern-recognition based approach to the problem of 
removal of polarized fringes from spectro-polarimetric data. 
We demonstrate that 2D Principal Component Analysis can be trained on 
a given spectro-polarimetric map in order to identify and
isolate fringe structures from the spectra. This allows us in
principle to reconstruct the data without the fringe component,
providing an effective and clean solution to the problem.
The results presented in this paper point in the direction of revising 
the way that science and calibration data should be planned for a typical
spectro-polarimetric observing run.
\end{abstract}

\maketitle

\section{Introduction} \label{sec:intro}

The observation and interpretation of wavelength dependent
polarization signals in spectral lines is the primary method
for the diagnostics of anisotropic processes in astrophysical plasmas,
such as those induced by the presence of deterministic electric and 
magnetic fields \cite[e.g.,][]{St94,LL04,CL08,TB10}, or by plasma collisions 
with collimated beams of ions \cite[e.g.,][]{Fu08}. At the same time,
the depolarizing effects by isotropic collisions and by quasi-random 
electro-magnetic fields can yield information on the
density of the plasma constituents, as well as important insights on 
the overall complexity of turbulent plasmas at diverse spatial and 
temporal scales \cite[e.g.,][]{Ca09b}.

The amplitudes of polarization signals observed in astrophysical 
plasmas vary widely, ranging from the very small signatures 
($\lesssim 10^{-3}\,I$, where $I$
is the radiation intensity) typical of the weak-field modifications of 
scattering polarization by the Hanle effect \cite[see, e.g.,][]{LL04}, 
to large amplitudes ($\gtrsim 10^{-1}\,I$) induced by the Zeeman 
effect in the presence of strong magnetic fields, such as those found 
in sunspots.
The weaker polarization signals are easily swamped by systematic errors
associated with instrumental effects, which are often difficult to model 
to the level of precision (\emph{polarization accuracy}) that is needed 
in order to isolate the true signals coming from the observed physical 
system. The calibration 
of this \emph{instrumentally induced polarization} is a difficult art. 
At the same time, the pursuit of ever finer spatial and temporal scales 
in the investigation of astrophysical plasmas puts continually growing 
demands on both sensitivity (i.e., signal-to-noise) and accuracy of 
polarimetric observations 
\cite[e.g.][]{Ri08}. Even higher demands are being made by the
scientifically critical need to measure chromospheric magnetic fields
\citep{Ju10}.

\begin{figure}[t!]
\centering
\includegraphics[height=2.9in]{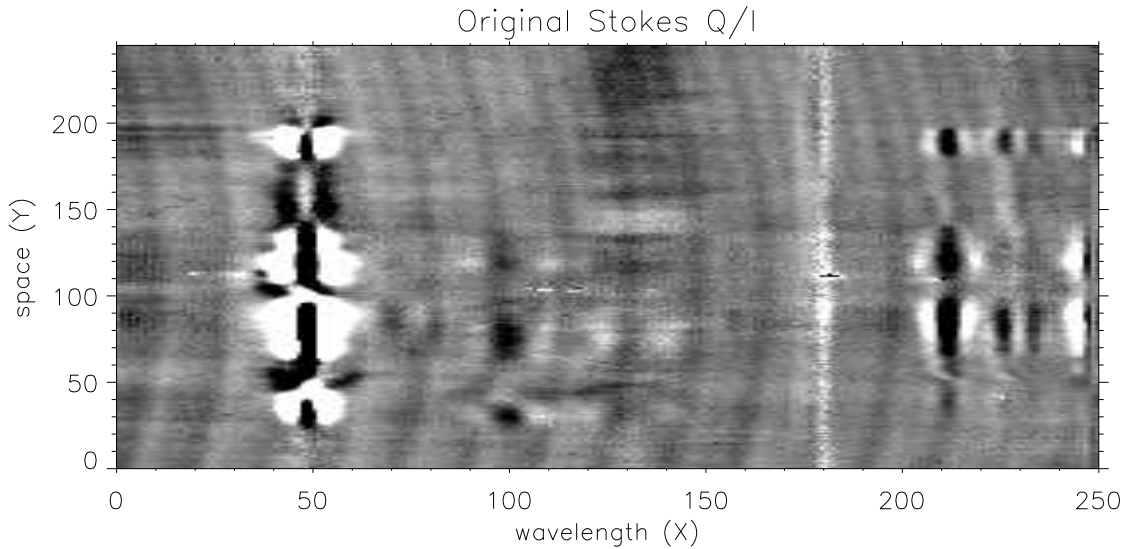}
\caption{\label{fig:example}
Example of spectro-polarimetric data, showing real spectral line 
polarization signatures superimposed to instrumental effects, including 
polarized fringes and detector noise. The abscissa is wavelength,
while the ordinate is position along the instrument's slit. The spectral
range includes the line of \ion{Si}{1} at 1082.7\,nm (around $X=50$), the
two components of \ion{He}{1} at 1083\,nm (around $X=100$ and $X=130$), the
telluric H$_2$O line (around $X=180$), and the \ion{Ca}{1} doublet at
1083.3\,nm (approximately from $X=210$ and $X=230$).}
\end{figure}

Polarized fringes are commonly found in spectro-polarimetric 
data. These are interference patterns that arise because of the presence 
of optical components (also including air) in a spectro-polarimeter, which 
have different refractive and/or birefringent properties. Such
components include polarization modulators, polarizing beam-splitters, 
and any optical system where parallel optical interfaces may occur
(e.g., interference filters, detector windows).
These 
fringes have the appearance of more or less regular bidimensional 
patterns, often curved, which typically unfold along the spectral 
dimension of the data (see Fig.~\ref{fig:example}). We refer to review 
studies of polarized fringes \citep{Se03,Cl04} for a thorough description 
of this phenomenon. Here, we are interested exclusively with the treatment of 
this artifact during data reduction.

The treatment of polarized fringes has been a recurring problem for 
the reduction and analysis of spectro-polarimetric data. The state of 
the art is to attempt removal of these (and other) instrumental effects
using Fourier methods or wavelet analysis \cite[e.g.,][]{RH06}. 
Fourier filtering has been successful at removing various types of
data artifacts, when their range of spectral and/or spatial frequencies 
is clearly separated from that of the actual signal of the observed 
source (L.~Kleint, private communication). 
Wavelet analysis attempts to identify the dominant frequencies and 
phases of the fringe pattern in a data frame, when the artifacts are not
strictly periodic. \cite{RH06} have developed 
a localized solution to fringe pattern reconstruction by employing 
two-dimensional wavelet transforms, which excel at tracking smooth variations in 
phase and amplitude of a periodic signal. The shape of a fringe pattern 
can be isolated in the wavelet space of individually transformed image 
rows, each row corresponding to a spatial point in the map. An inverse 
wavelet transform of the fringe space then can reconstruct the fringe 
pattern in the spatial domain. This can be a particularly powerful 
method for removing fringes in flat field images, but its application 
to object images (especially polarized spectral images) is complicated 
by the contribution of the targeted signal to the local wavelet transform.


Unfortunately, fringe patterns are seldom regular, having an intrinsic 
bidimensional structure, with variations of the amplitude, frequency, 
and phase, which can be significant in both the spectral and spatial 
dimensions (see Fig.~\ref{fig:example}). Often fringe patterns result 
from the combination of more than one component, and this combination 
can also vary smoothly across the dataset. In this case, a
two-dimensional wavelet 
analysis of the observations is challenged by the need to treat every 
frame separately. In this paper, we propose instead that the identification 
and removal of polarized fringes might be better approached as a problem of 
pattern recognition.

Since polarized fringes typically arise within the
spectro-polarimeter,\footnote{A notable exception is represented by
fringing that is caused by the polarization calibration optics, which
may reside outside the spectro-polarimeter, often well upstream in 
the optical system of the telescope.}
we expect their structure to be predominantly a function of the
instrument configuration. Let us consider a scanning slit instrument.
During the spatial scan of a given 
target, the instrument configuration is approximately fixed. 
We can expect that the polarized fringes will constitute an 
approximately time-independent pattern within that particular 
dataset, although there may be some dependence of the fringes' 
appearance (in both amplitude and phase) on the polarization state 
of the light entering the spectro-polarimeter. 
As the spatial scan is acquired, the polarimetric signal of the target 
will then change over an approximately constant fringe background. 
Heuristically, an ``orthogonality'' exists between the true 
polarimetric signal that we wish to analyze and this ``fixed'' fringe 
background, as a consequence of the fact that the two sources of the
line signals and of the background are largely uncorrelated. This 
suggests that the problem of identification and 
removal of polarized fringes should be approached as a problem of 
pattern recognition and feature extraction from a two-dimensional
dataset. For this paper we decided to approach this problem using 
two-dimensional Principal Component Analysis \cite[2D PCA;][]{Ya04}. 
Other methods could potentially be adopted instead, which also have 
been used for the separation of signals from uncorrelated sources, 
such as the Independent Component Analysis \cite[ICA; e.g.,][]{JH91}. We 
defer the study of some of these alternative methods for the problem of
the identification and removal of polarized fringes to future work.

The simple fact that spectral line signals and polarized fringe background
are largely uncorrelated is central to the success of the PCA approach to
the problem at hand. This concept is nicely summarized by \cite{Jo02} in the
Introduction to his book: ``\emph{The central idea of principal component
analysis (PCA) is to reduce the dimensionality of a data set consisting
of a large number of interrelated variables, while retaining as much as
possible of the variation present in the data set. This is achieved by
transforming to a new set of variables, the principal components (PCs),
which are uncorrelated, and which are ordered so that the first
\textbf{few} retain most of the variation present in \textbf{all} of the
original variables.}''

Preliminary considerations on the problematics associated with 
PCA filtering of polarized fringes from Stokes profiles were already 
advanced in a study on \emph{compressed sensing} of experimental data 
\citep{AL10}.
In this work, we provide a novel and more in-depth investigation of the problem.
In Sect.~\ref{sec:PCA} we summarize the main ideas behind two-dimensional 
PCA, and show how these can be applied to the specific problem of the
removal of polarized fringes from spectro-polarimetric data. 
In Sect.~\ref{sec:data}, we describe the observations from which the 
datasets used for the testing of the method were extracted. 
In Sect.~\ref{sec:results} we present various examples of application 
of the method to the datasets previously described, and comment on the 
quality and reproducibility of the results. 
Finally, in our concluding remarks, we briefly discuss possible 
observation and data calibration strategies that could help fully
realize the potential of the proposed method.

\section{Two-dimensional Principal Component Analysis}
\label{sec:PCA}

The theory behind PCA is well
established \citep{Pe01,Jo02}. PCA has been successfully
applied to a related but different problem of the inversion of 
spectro-polarimetric data for the inference 
of the thermodynamic and magnetic structure of the solar atmosphere 
\citep{Re00,SN01,LC02,Ca05,Ca09a}. In this application, PCA is used 
to identify an
orthogonal set of spectral \emph{eigenfeatures} characteristic of a 
given line formation model in a magnetized atmosphere. One then 
determines the principal components (projections) of the observed spectra 
on this orthogonal eigenbasis, and searches within a pre-calculated 
database of model profiles for the closest set of components to the 
observed set. The quality of the inversion is estimated by the 
``PCA distance'' between the observed and inverted points in the 
PCA-component space. This distance is akin to the usual $\chi^2$ 
estimator for a non-linear least-squares fit. In this special application, 
PCA then produces a best fit of the observed Stokes profiles as a 
function of wavelength. PCA is able to capture the structures of the
observed profiles that correspond to the physical model used for the
construction of the inversion database. Systematic errors introduced by
the instrument, or by physical mechanisms not included
in the model, cannot be captured by the PCA inversion, and consequently
they tend to increase the inversion noise (i.e., the PCA distance).
Because every spatial point is emitting incoherently from all the other
points in the map, the spectro-polarimetric inversion must be performed
separately for each spatial point of the scan.

The application of PCA to our problem is instead subject to rather 
different constraints.
First of all, polarized fringes are typically very hard to model, and 
their appearance -- such as frequency, phase, as well as shape -- can 
be very specific of the particular observing and instrumental setup, 
and may change significantly between different observations. Therefore, 
rather than creating a model-based database of fringes to compare with 
the observations, one should extract the PCA eigenfeatures of the 
fringe pattern directly from the observed data.
Another substantial difference is that the treatment of fringes is
better done simultaneously on the entire detector frame, because of 
the intrinsic bidimensional nature of these instrumental artifacts. 
The correct approach should then be analogous to the one adopted in the 
application of PCA to face recognition.

To further proceed, we need to choose one among the various PCA algorithms 
that have been developed for face recognition \citep{KS90,TP91,Ya04}.
The 2D PCA algorithm proposed by \cite{Ya04} appears to outperform other
approaches, and so we selected it for our problem. We summarize here 
below the fundamental ideas of this algorithm.

We indicate with $\tens{A}_i$, for $i=1,\ldots,N$, 
a set of images (e.g., the sequential series of the frames in a 
spectro-polarimetric map), each represented by a $m\times n$ matrix. 
From these, we can create the $n\times n$ covariance matrix
\begin{equation} \label{eq:covariance}
\tens{C}=\frac{1}{N}\sum_{i=1}^N (\tens{A}_i-\bar\tens{A})^T
(\tens{A}_i-\bar\tens{A})\;,\qquad
\bar\tens{A}=\frac{1}{N}\sum_{i=1}^N \tens{A}_i\;.
\end{equation}
It is demonstrated \cite[e.g.,][]{Jo02} that the \emph{optimal} set of
projection vectors for the decomposition of the data, for the purpose of
extracting its \emph{principal components}, is represented by the $n$ 
eigenvectors of the covariance matrix. These eigenvectors can 
conveniently be determined by performing the \emph{singular value 
decomposition} (SVD) of $\tens{C}$. 
The result is a set of mutually orthogonal, $n$-dimensional vectors, 
$\bm{U}_j$, with $j=1,\ldots,n$, which can be interpreted as the column 
vectors of a $n\times n$ orthogonal matrix 
$\tens{U}=\{\bm{U}_1,\ldots,\bm{U}_n\}$. It is customary to order the 
set of eigenvectors $\{\bm{U}_i\}_{i=1,\ldots,n}$ (the
\emph{eigenfeatures}) according to the decreasing amplitude of the 
associated singular values, $\sigma_i$, so that $\bm{U}_1$ corresponds 
to the singular value with the largest amplitude. In turn, the singular 
values can be interpreted as the 
weights of the corresponding eigenfeatures in their contribution to 
the dataset $\{\tens{A}_i\}_{i=1,\ldots,N}$. The 
set $\{\bm{U}_i\}_{i=1,\ldots,n}$ is complete, and thus forms a basis for
the dataset $\{\tens{A}_i\}_{i=1,\ldots,N}$.

Once the basis $\{\bm{U}_i\}_{i=1,\ldots,n}$ has been
determined, the set of \emph{principal components} (or projections) 
for each of the elements of the dataset $\{\tens{A}_i\}_{i=1,\ldots,N}$
is
\begin{equation} \label{eq:component}
\tens{V}_i=\tens{A}_i\tens{U}\;,\qquad i=1,\ldots,N\;.
\end{equation}
We note that each $\tens{V}_i$ is a $m\times n$ matrix, which can be
thought of as the set of $m$-dimensional column vectors, 
$\{\bm{V}_{i,j}\}_{j=1,\ldots,n}$, such that
\begin{equation} \label{eq:subcomponesn}
\bm{V}_{i,j}=\tens{A}_i\bm{U}_j\;,\qquad 
	i=1,\ldots,N\;,\quad j=1,\ldots,n\;.
\end{equation}
Since $\tens{U}$ is an orthogonal matrix with unit determinant, 
eq.~(\ref{eq:component}) can be immediately inverted,
\begin{equation} \label{eq:reconstruction}
\tens{A}_i=\tens{V}_i\tens{U}^T=\sum_{j=1}^n \bm{V}_{i,j}\bm{U}_j^T\;,
	\qquad i=1,\ldots,N\;.
\end{equation}
This equation represents the algorithm for the reconstruction of the
image $\tens{A}_i$ from its PCA component matrix, $\tens{V}_i$, and the
PCA eigenbasis $\tens{U}$. In particular, any truncation of the
summation on the left of eq.~(\ref{eq:reconstruction}), up to some 
index $j=j_{\rm max}<n$, will provide
an approximate reconstruction of the original image,
\begin{equation} \label{eq:reconstruction.partial}
\tilde\tens{A}_i=\sum_{j=1}^{j_{\rm max}} \bm{V}_{i,j}\bm{U}_j^T\;.
\end{equation}

If the image set is characterized by a low degree of entropy (that is, 
a high level of ordering -- like for the image of an object, or, 
in our case, the Sun's line spectrum between two fixed wavelengths), 
the amplitudes of the singular values drop very fast -- typically, 
by several orders of magnitude within the first few eigenfeatures. This
implies that the information content of the data is practically confined 
within a set of eigenfeatures which is much smaller than the complete
set,\footnote{Incidentally, this property provides a convenient means
for data compression.} which, on the other hand, will also contain 
information on less significant and/or more random features of the 
images, including noise.

Equation~(\ref{eq:covariance}) shows that a perfectly constant background 
affecting all images $\tens{A}_i$ is completely removed from the
expression of the covariance matrix $\tens{C}$. In such an ideal case, 
the eigenvectors $\{\bm{U}_i\}_{i=1,\ldots,n}$ constitute an optimal 
set of projection vectors for the relevant signal in the data, but not 
for the background. One should then expect to be able to reconstruct the
signal to any degree of precision (including noise), 
while leaving the constant background out. This, of course, is never
the case with real spectro-polarimetric data. Residual fringing is
going to be present in the covariance matrix, and this ultimately
affects the ability to reconstruct spectral signals that happen to be
co-located with the fringes, and with comparable amplitudes.

Note that the form of eq.~(\ref{eq:covariance}) implies that the
bidimensional dataset is summed (\emph{contracted}) over the $m$ rows 
of the differential images $(\tens{A}_i-\bar\tens{A})$. Alternatively,
we could have defined the covariance matrix as the average of the
products $(\tens{A}_i-\bar\tens{A})(\tens{A}_i-\bar\tens{A})^T$
over the dataset, implying a contraction over the $n$ columns of the 
differential images. In PCA applications to face 
recognition it typically is irrelevant whether the 
PCA covariance matrix is computed by contracting the bidimensional
data over the $X$- or the $Y$-axis. For the analysis of Stokes maps,
instead, the spectral dimension (conventionally identified in this work
with the $X$-axis) represents a privileged coordinate. The reason is
two-fold. First of all, it is essential for our problem that we preserve
the distinction among the different spectral features in a Stokes map.
In fact, because different spectral lines may be formed under very diverse 
atmospheric conditions, and have distinct thermal and magnetic diagnostic
properties, it is important that their characteristics be kept distinct
in the PCA decomposition of the Stokes data. If we were instead to 
contract the data along the spectral dimension, the diagnostic information 
from all spectral lines would be merged together. Furthermore, preserving 
the spectral dimension in the PCA decomposition of a Stokes map allows us 
to isolate a specific subinterval of the spectral domain, if needed. The 
second argument is that polarized fringes tend to occur preferentially 
along the spectral dimension. Again, if we were to contract the data over
wavelength, the spectral information of the polarized fringes would
be completely merged with that from the spectral lines. The immediate
consequence is that each of the PCA eigenfeatures would then always
contain the information from both spectral lines and polarized
fringes, making the filtering out of fringes impossible. All examples 
shown in the following discussion rely on the preservation of the 
spectral information of the Stokes map in the PCA decomposition. For
this reason, we adopt the definition of the covariance matrix given by
eq.~(\ref{eq:covariance}).

\section{Observations} \label{sec:data}

Spectro-polarimetric observations of an active region near the solar
limb were taken on September 22, 2011 with the Facility Infra-Red
Spectro-polarimeter \cite[FIRS;][]{Ja10} and the Interferometric
BIdimensional Spectrometer \cite[IBIS;][]{Ca06,RC08}. Both instruments 
are deployed at the Dunn Solar Telescope (DST) of the National Solar 
Observatory on
Sacramento Peak (NSO/SP, Sunspot, NM). FIRS was used in a single-slit, 
dual-beam mode, with the slit oriented tangentially to the solar limb. 
The $75\arcsec$-long projected slit was scanned across the solar 
image, in 70 steps of $0.65\arcsec$,  to produce images in four 
polarimetrically modulated states, $S_i$. The spectral range of the
observations spanned from 1081.93 to 1085.01\,nm.

The data were reduced using software originally developed by 
S.\ Jaeggli \citep{Ja11} and modified by one of the authors (TS). 
The data reduction followed the standard procedures: 1) correction for 
non-linearities of the detector; 2) subtraction of dark frames;
3) division by flat fields; 4) co-registration of the two beams,
including corrections for image rotation; 5) polarization calibration;
6) de-modulation of the signals $S_i$ to convert them into the 
corresponding Stokes parameters $I$, $Q$, $U$, and $V$. Special care 
was taken to acquire flat fields before and after the scans analyzed
here, which were then linearly interpolated in time before being applied 
to the science data. This flat-field correction eliminated the
dominant part of the signal contributed by the polarized fringes. 
However, there remained residual fringes and some other detector 
artifacts in the processed data. The method proposed here for the
identification and removal of fringes has been applied to these reduced
data.

\begin{figure}[t!]
\centering
\includegraphics[height=4.8in]{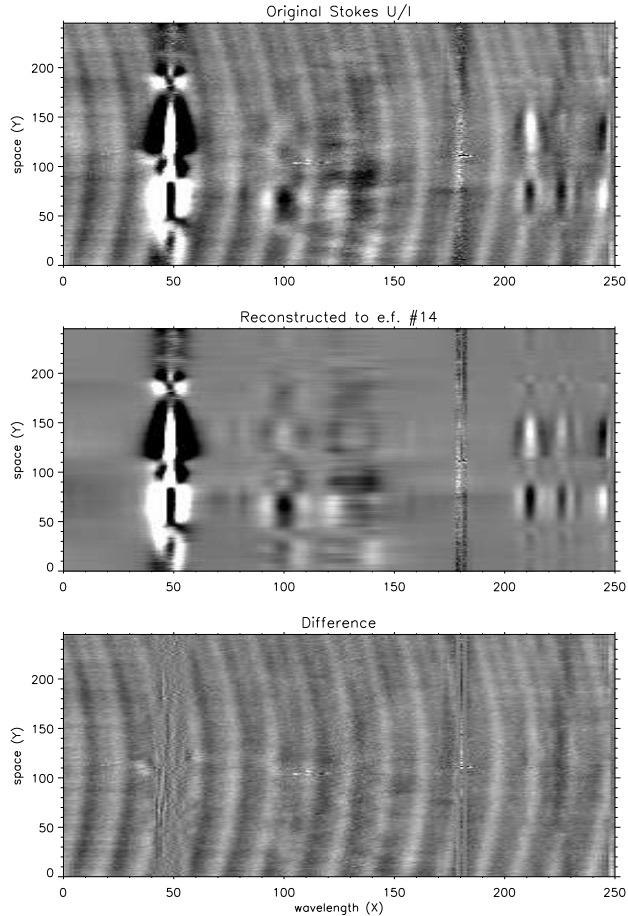}
\caption{\label{fig:StokesU.full}
Example of PCA filtering of polarized fringes from a Stokes-$U$
spectrum in the wavelength region of \ion{He}{1} 1083\,nm. 
The \ion{He}{1} triplet spans approximately between $X=90$ and 
$X=150$ on the horizontal axis. The top panel shows the 
original data, the middle panel the PCA reconstructed data (using 
the first 14 PCA eigenfeatures), and the bottom panel shows the 
difference of the two. We see how most of the large-scale fringes have 
been taken out in the PCA reconstruction, revealing more clearly ``real''
solar spectral features.}
\end{figure}

\begin{figure}[p!]
\centering
\includegraphics[height=4.8in]{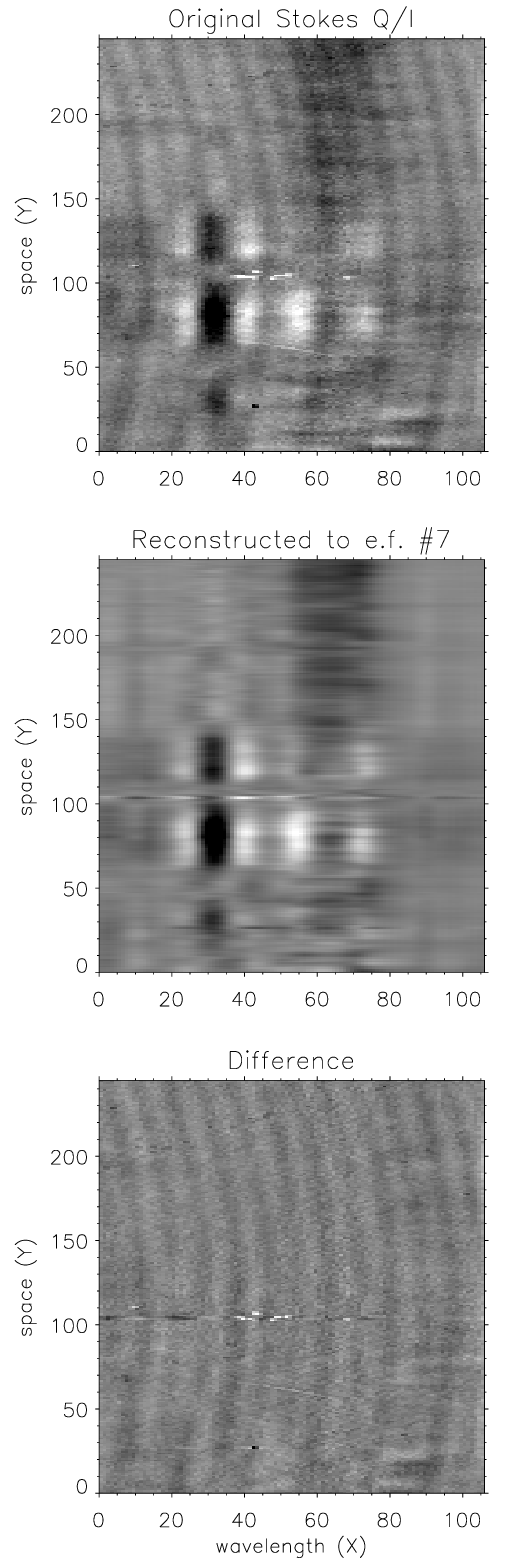}
\includegraphics[height=4.8in]{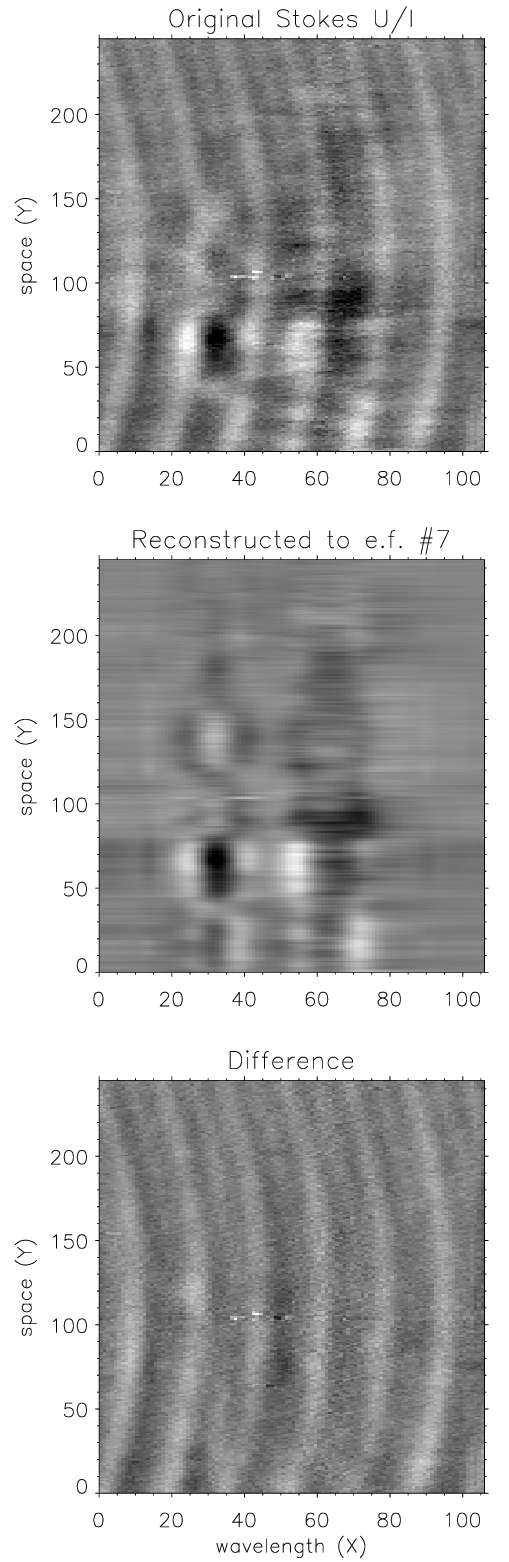}
\includegraphics[height=4.8in]{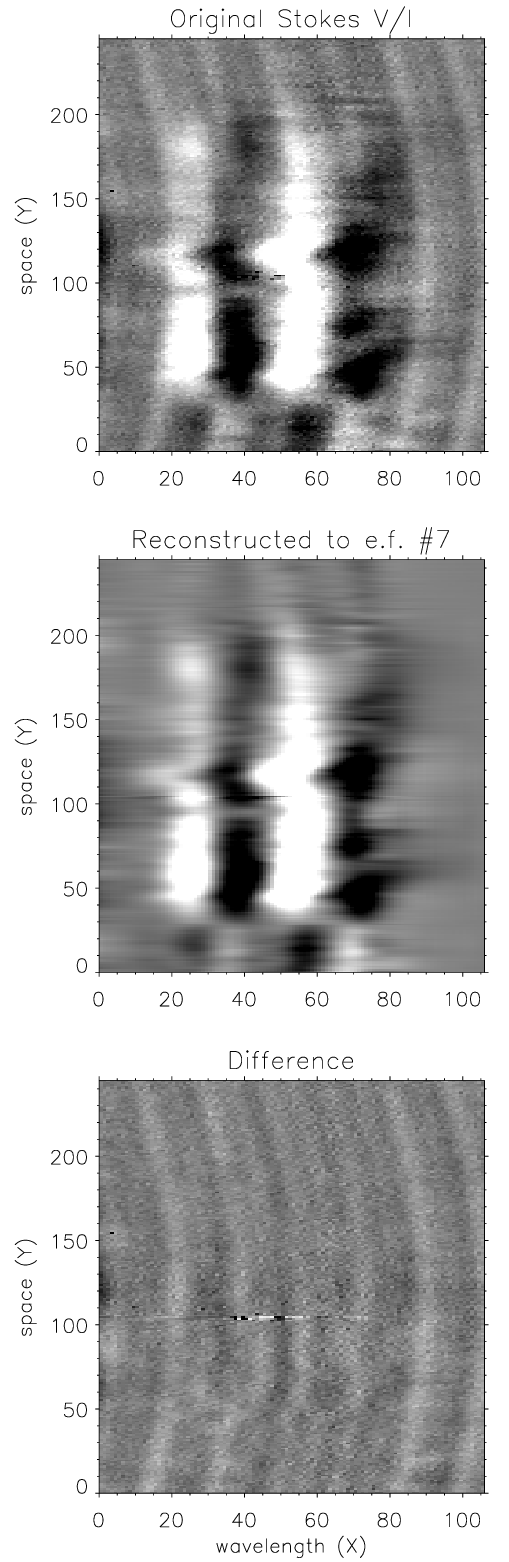}\vspace{9pt}
\includegraphics[width=.325\hsize]{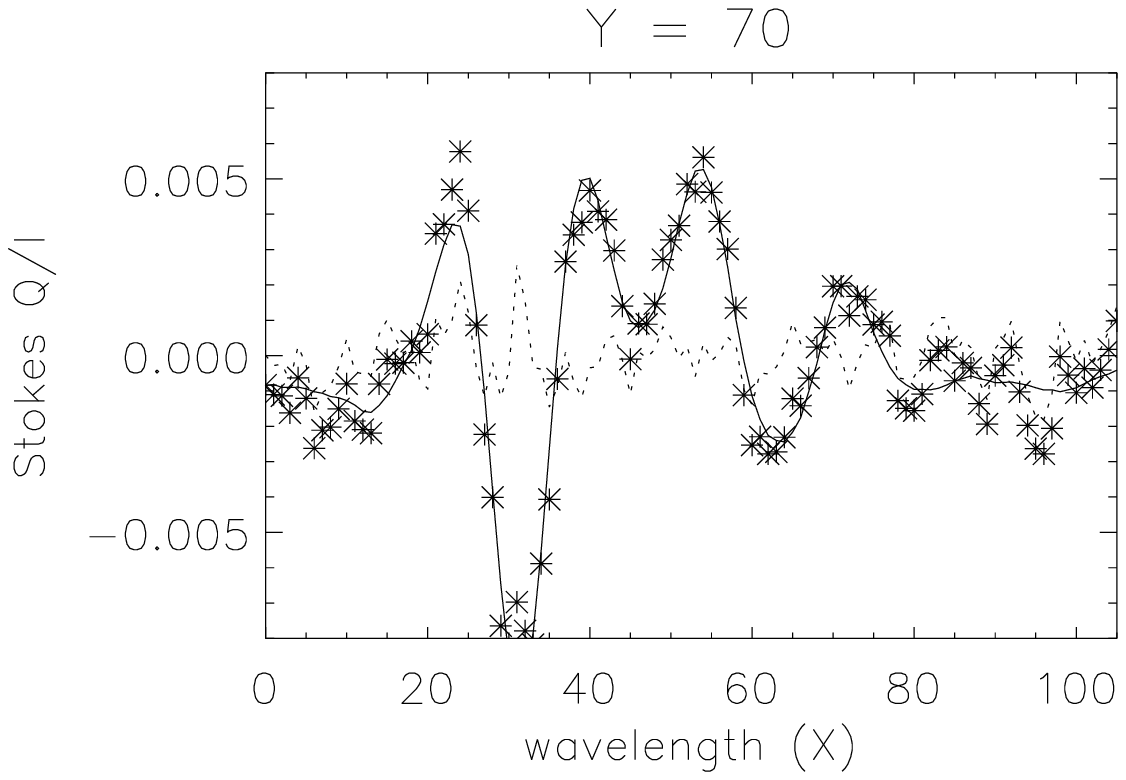}
\includegraphics[width=.325\hsize]{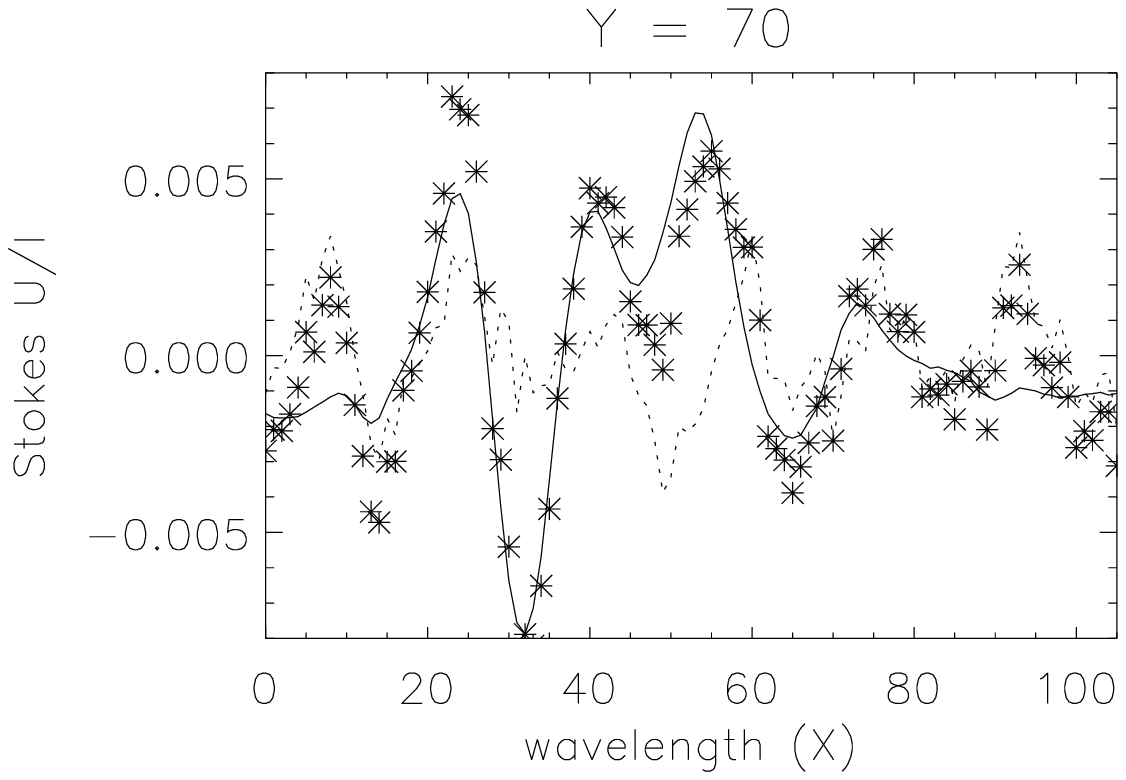}
\includegraphics[width=.325\hsize]{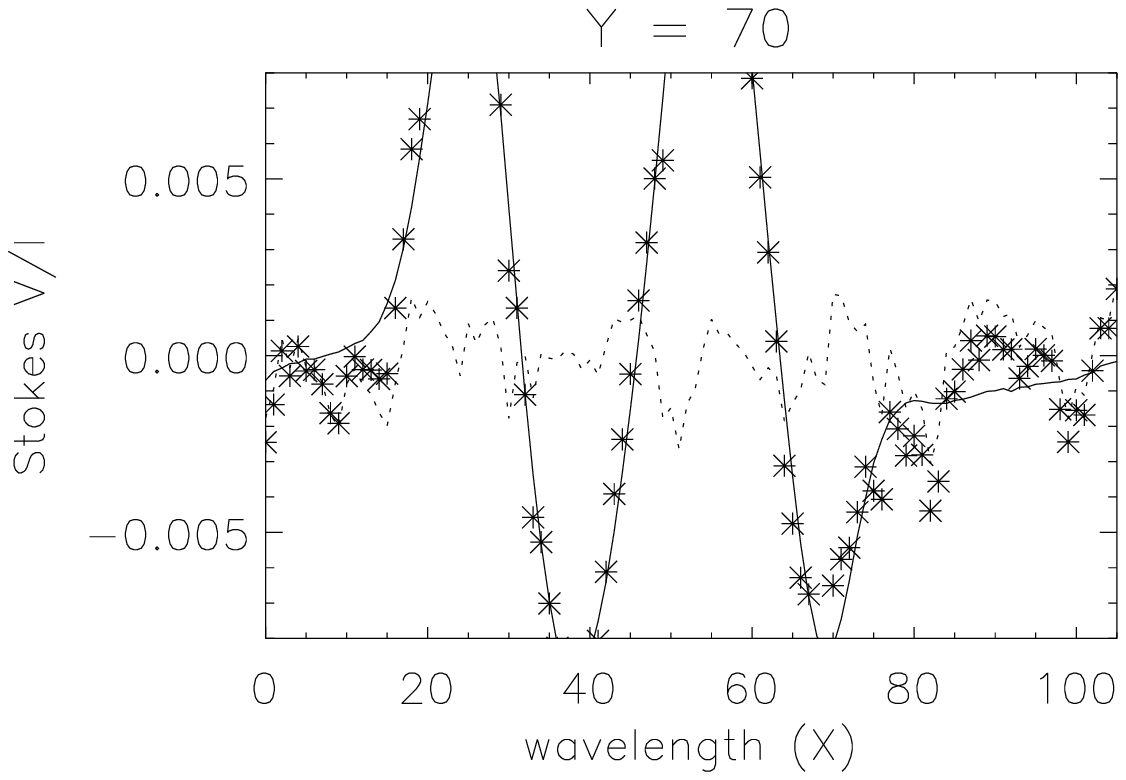}
\caption{\label{fig:StokesAll.HeI}
Same dataset as in Fig.~\ref{fig:StokesU.full}, but with the PCA
filtering applied only to the spectral region encompassing \ion{He}{1}
1083\,nm. The avoidance of the strong signatures of the \ion{Si}{1} line 
and the H$_2$O atmospheric line allows for an efficient removal of the 
fringe pattern already with half the number of eigenfeatures as used for
Fig.~\ref{fig:StokesU.full}. The bottom panels show the Stokes profiles 
for the spatial point $Y=70$. Star symbols show the data, the
continuous line shows the reconstructed data, and the dotted line the
difference of the two.}
\end{figure}

\section{2D PCA of Stokes spectra with polarized fringes}
\label{sec:results}

As we anticipated in both Sects.~\ref{sec:intro} and \ref{sec:PCA}, 
the ability of PCA to isolate a background pattern within a dataset 
depends critically on how constant the appearance of the background 
pattern is throughout the dataset. In particular, to eliminate any 
unwanted artifact from the scientifically relevant data, it is 
important that the realizations of the relevant signals throughout 
the dataset be as varied as possible (and ideally vanish for a 
\emph{statistically significant} number of frames), while 
the pattern of the artifact should remain practically constant. This 
allows PCA to ``recognize'' the signals and the background pattern 
as originating from uncorrelated sources, and ultimately to isolate 
the pattern into an ``orthogonal'' subspace with respect to the 
signals. \emph{This is the fundamental property enabling a PCA 
reconstruction of the scientific data that excludes the artifact.}

An application of the concepts presented so far is illustrated by 
Fig.~\ref{fig:StokesU.full}, showing the Stokes-$U$ spectrum 
in the wavelength region of \ion{He}{1} 1083\,nm, taken from the
observations presented in Sect.~\ref{sec:data}. This 
particular spatial scan encompasses a strongly magnetized active region 
with an overlying filament, as well as a portion of the quiet Sun. 
In these plots, the
spectral dispersion (wavelength) is along the horizontal axis, and the
spatial dimension is along the vertical axis. From left to right, we
recognize the spectral features of the \ion{Si}{1} line at 1082.7\,nm,
the \ion{He}{1} triplet at 1083\,nm, an H$_2$O telluric absorption line, 
and the \ion{Ca}{1} doublet at 1083.3\,nm (the leftmost signal is part 
of the \ion{Na}{1} triplet at 1083.5\,nm).

The \ion{Si}{1} line shows a dominant signal in all frames of the map, 
with an average central depth of about 0.5 in intensity, and 
polarization levels typically between 10 and 20\%. The H$_2$O telluric 
line also shows some polarization, at the level of 1\% for linear 
polarization and a few times smaller for circular
polarization. This must be an instrumental artifact, indicating 
a problem with the demodulation of the polarized signals into the Stokes 
vector, likely due to incorrect flat-fielding caused by the 
non-linearities of the FIRS IR detector \citep{Ja11}. Because of this, 
the H$_2$O line also appears in all 
the frames of the Stokes map. The polarized fringes are also 
visible as curved structures spanning the wavelength domain. For the 
example of Fig.~\ref{fig:StokesU.full}, we 
performed a PCA decomposition of the Stokes-$U$ map, consisting 
of 45 slit positions over the solar disk, and reconstructed the data 
for one particular slit position, using only the first 14 eigenfeatures 
out of the 250 total
(in our decomposition, the number of eigenfeatures equals the number of
wavelength points in the scan; see Sect.~\ref{sec:PCA}). We note how the 
signals are well
captured by the PCA reconstruction, while the fringes are ``confined'' 
within the orthogonal subspace corresponding to eigenfeatures larger 
than \#14, almost everywhere in the spectral domain of the map. The 
spectral ranges of the \ion{Si}{1} and H$_2$O lines are an exception, 
precisely because of their strong polarization signals appearing in all 
frames of the map, which prevents the PCA decomposition from separating 
these signals from the background fringes. In other words, at these
particular wavelengths, PCA is unable to conclude that the spectral line 
signal and the fringes are uncorrelated.

\begin{figure}[t!]
\centering
\includegraphics[width=.495\hsize]{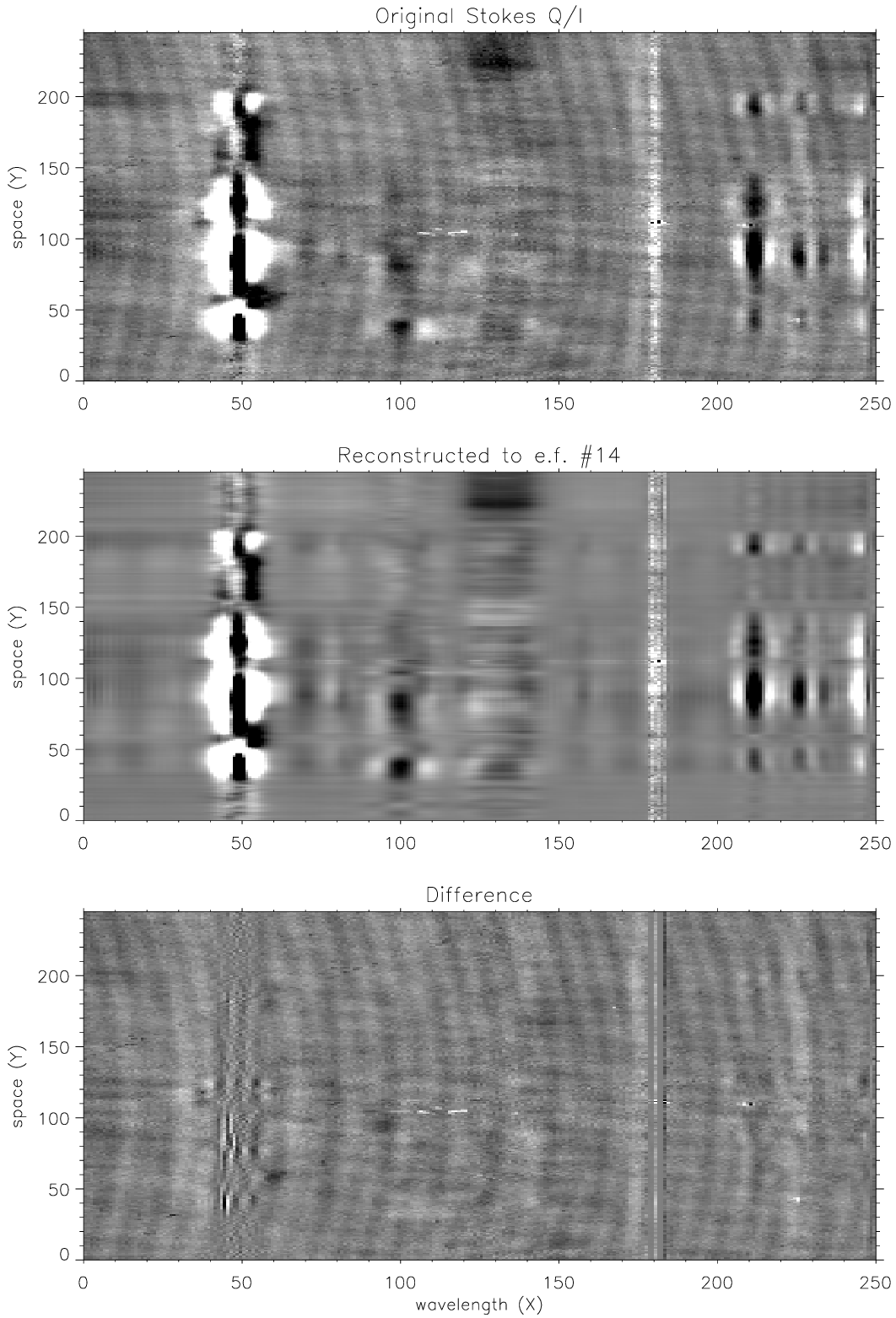}
\includegraphics[width=.495\hsize]{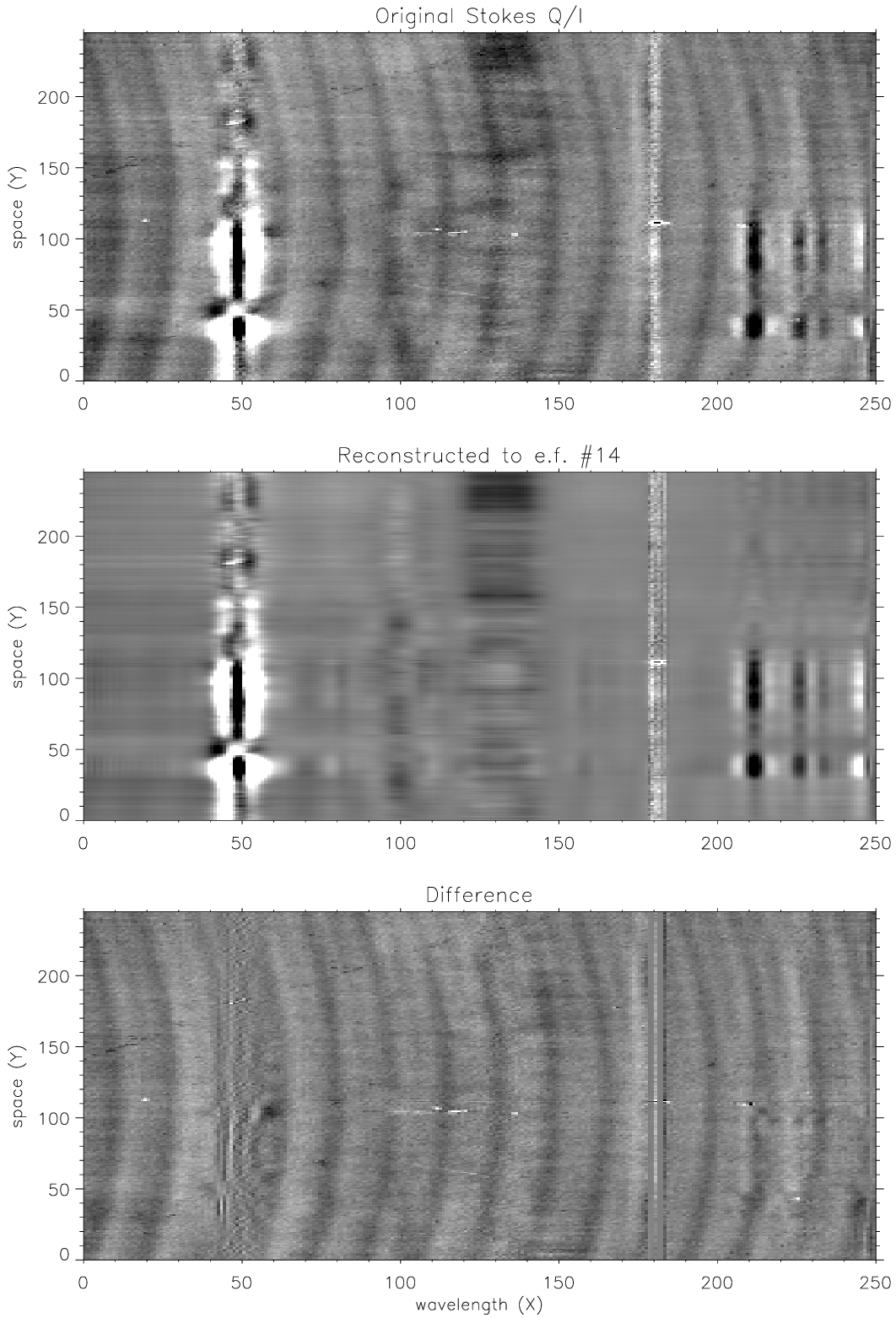}
\caption{\label{fig:twofreq}
PCA reconstruction of two different Stokes-$Q$ scan steps from the same
dataset, showing how the fringe pattern can appear with distinct 
frequencies at different positions in the map. The PCA decomposition in
this case is able to identify the two frequencies in the dataset, and 
subtract the proper combination of the respective contributions.}
\end{figure}

\begin{figure}[t!]
\centering
\includegraphics[width=.95\hsize]{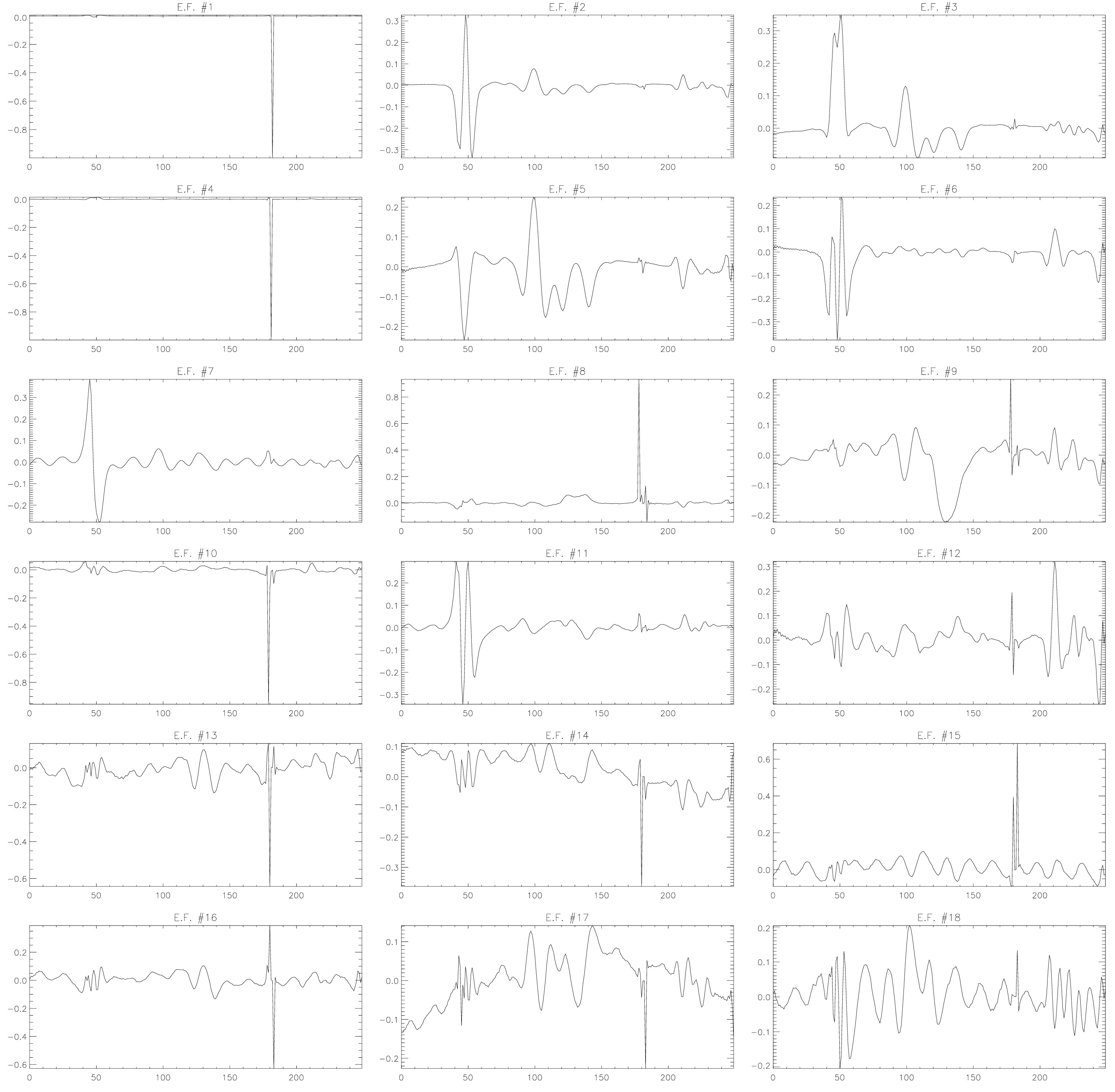}
\caption{\label{fig:eigenbasis}
The first 18 basis eigenfeatures $\bm{U}_i$ for the Stokes-$Q$ dataset 
from which the two scan steps of Fig.~\ref{fig:twofreq} were taken.}
\end{figure}

Figure~\ref{fig:StokesAll.HeI} shows instead the results of the PCA
decomposition applied to a restricted set of wavelengths, encompassing
just the \ion{He}{1} triplet. The three columns show the 
intensity-normalized $Q$, $U$, and $V$ Stokes parameters. Like in the 
previous example, PCA decomposition is able to isolate the
fringes, because of the much more diverse appearance of 
the \ion{He}{1} polarization signals in the map. However, we also note
how the fringe background can be isolated and removed already using a 
PCA reconstruction that takes into account only half of the eigenfeatures 
that were used for Fig.~\ref{fig:StokesU.full}. This is not surprising,
since the low-order PCA eigenfeatures are always dominated by the 
structures that produce the largest variance in the data, and thus 
the strong signals of the \ion{Si}{1} line, as well as the sharp 
H$_2$O line, occupy exclusively the first two eigenfeatures in the PCA 
decomposition of the full-range spectral maps depicted in 
Fig.~\ref{fig:StokesU.full}, while
the weakest signals of the \ion{He}{1} triplet and the polarized
fringes only show up in later orders. In the case of the spectral maps
of Fig.~\ref{fig:StokesAll.HeI}, instead, the variance of the data is
dominated by the \ion{He}{1} signals, which therefore appear in the PCA 
decomposition already in the lowest order. In fact, increasing the 
number of eigenfeatures in the reconstruction of the examples of 
Figs.~\ref{fig:StokesU.full} and \ref{fig:StokesAll.HeI} does not 
improve the quality of the 
fringe removal, but rather the fringe pattern starts leaking back 
into the reconstructed image. Again, this is due to the imperfect
subtraction of the fringe background in the creation of the covariance
matrix of the data.


In the datasets that we have analyzed from the FIRS instrument, the
fringe patterns of a spectro-polarimetric map seem to occur always with 
one frequency and its first harmonic, as well as with various 
combinations of the two, although seemingly with nearly constant phases 
throughout the map. Apparently, PCA is able to identify the right 
combinations of these two basic fringe patterns and remove them from 
the spectral line signal (see Fig.~\ref{fig:twofreq}). As long as a 
statistically significant number of samples of the recurring 
frequencies and phases of the fringes are present in a given 
observation, without contamination from the data (i.e., with vanishing
spectral signals), we can reasonably assume -- and is demonstrated in 
practice -- that the PCA decomposition will be able to identify those 
characteristics, and their contributions to any given frame of the map. 
On the other hand, if the frequency and/or phase of polarized fringes vary 
continually with the map step of a given dataset, then we can expect
that the removal of fringes by PCA cannot be successfully accomplished
within that individual dataset.

%

We conclude this section by commenting on possible 
strategies to decide the order of truncation of the PCA reconstruction 
that is needed for a specific dataset, since ideally one would like to
devise some type of automated procedure for fringe removal from Stokes data. 

The data that we have analyzed in this work were acquired during 
observations that were not specifically prepared to optimize the 
removal of polarized fringes. 
Therefore, the information content of fringes varies greatly between 
different datasets. As we have demonstrated, this affects both the 
quality of the fringe removal, as well as the number of PCA 
eigenfeatures that must be retained in order to attain the best 
compromise between fringe removal and preservation of the original 
spectral line signals. Because of this, there is some degree of 
\emph{subjectivity} in the truncation of the PCA eigenfeature expansion. 
We have already shown how clipping out strong signals from the spectral 
range that is to be analyzed in terms of principal components allows 
to retain a smaller number of expansion orders.

On the other hand, in the data presented here, the spectral information 
of the Stokes line profiles is never completely ``orthogonal'' to that 
of the polarized fringes. In addition, the amplitude of the 
fringes is typically of the same order of magnitude as the Stokes signals 
of the \ion{He}{1} triplet that we want to refine. Thus, many of the 
eigenfeatures that contribute to the fringe pattern across a dataset 
will also contain information on the line signals (such as velocity shifts
and asymmetries), which may be critical for the science. Under these 
conditions, the truncation of the PCA reconstruction expansion still 
requires human oversight, unfortunately, and so can hardly be automated.


\begin{figure}[t!]
\centering
\includegraphics[width=.495\hsize]{D103517_Sc20_Stk1_EV14.eps}
\includegraphics[width=.495\hsize]{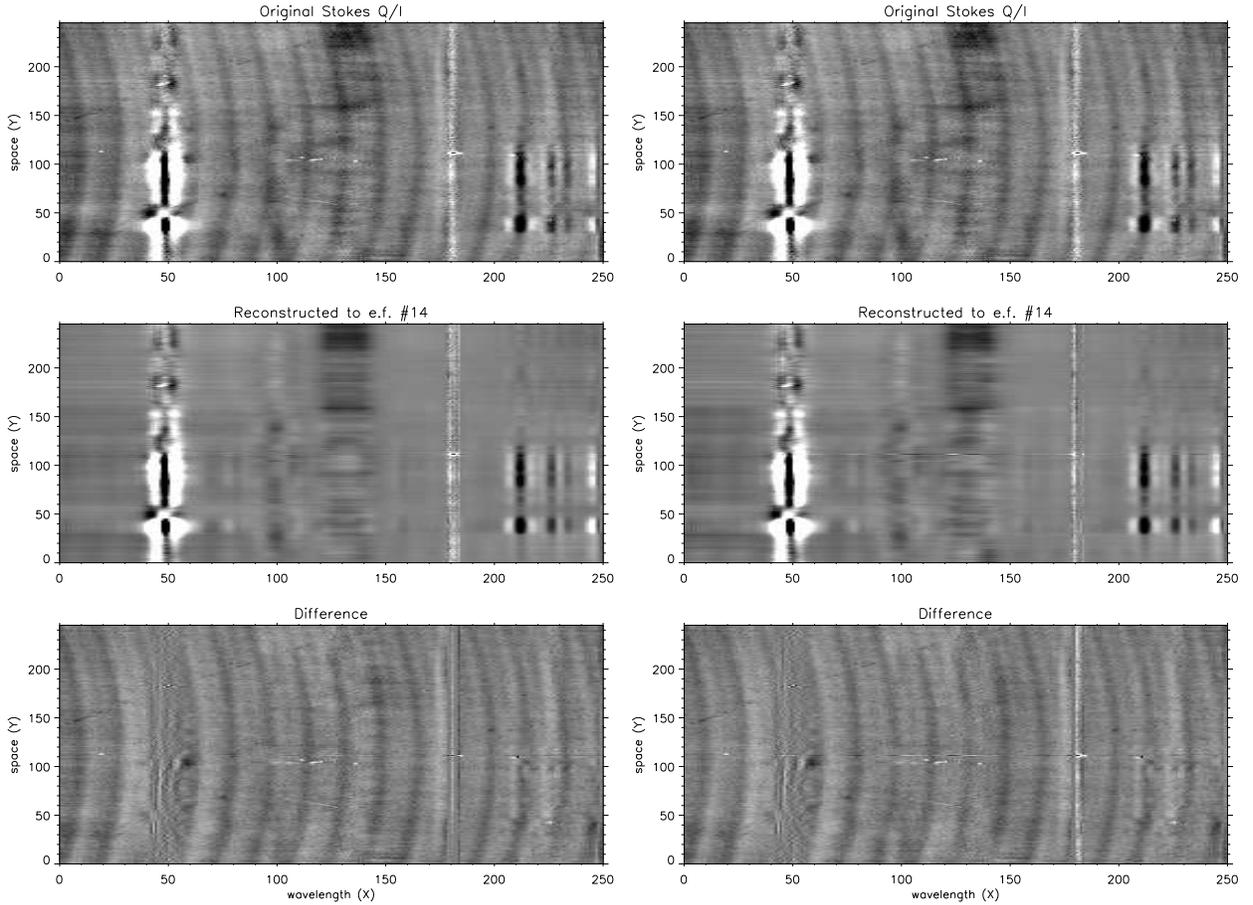}
\caption{\label{fig:ef_select}
Different PCA reconstructions of the same Stokes-$Q$ scan step; 
\emph{left:} using the full set of the first 14 eigenfeatures; 
\emph{right:} reconstructing the data by dropping the subset $\{1,4,14\}$
of eigenfeatures, and adding instead the subset $\{16,18\}$. We note how
the added eigenfeatures improve the reconstruction of the \ion{He}{1}
spectral data, while the quality of the fringe removal is preserved.}
\end{figure}

Beside truncation of the PCA expansion in data reconstruction, a
related issue is that of order selection, that is, the retaining or
dropping of critical eigenfeatures that work in favor of the 
preservation of the line signal simultaneously with the removal of 
polarized fringes and other artifacts. A cursory look at the 
structure of the basis eigenfeatures $\bm{U}_i$ and of the \emph{subimages}
$\bm{V}_{\ast,i}\bm{U}_i^T$ that contribute to a given scan step of the
dataset (cf.\ eq.~[\ref{eq:reconstruction}]) is often very
helpful to determine which order contains relevant spectral information, and
which one instead can be dropped in order to remove the fringe
background without critical loss of science data.

To illustrate this idea, Fig.~\ref{fig:eigenbasis} shows the first 18 
eigenfeatures for the dataset from which the two scan steps of
Fig.~\ref{fig:twofreq} were taken. We see at once some notable 
characteristics of this basis subset. The eigenfeatures \#1 and \#4 are 
completely dominated by the strong, sharp spectral signature of the 
H$_2$O line, while their contribution to the \ion{He}{1} spectral range 
is effectively zero.
The polarized fringes start appearing already in the eigenfeature \#7,
overlapped with a strong antisymmetric (i.e., velocity-type) component 
of the \ion{Si}{1} line. Evidently the eigenfeature \#15 is also 
completely dominated by fringes in the \ion{He}{1} region. The 
eigenfeatures \#14 and \#17 do not tend to zero at the boundary of the
spectral range. Likely this is caused by a residual gradient over the 
detector's frame after flat-fielding, which could affect a proper
collocation of the continuum level of the spectrum. With the exception
of \#7, the first eight eigenfeatures of Fig.~\ref{fig:eigenbasis} 
appear to be completely clear of the fringe background, only 
contributing to the line spectral signal that is relevant for science. 
All the successive eigenfeatures show instead some mixing of line
spectra and fringes. Elimination of these eigenfeatures will inevitably
determine some loss of science data, although this may occur at a level 
which lies below the target sensitivity of a particular observation.

Figure~\ref{fig:ef_select} illustrates an application of this strategy of 
order selection. The left side shows the same Stokes-$Q$ scan step used 
also for Fig.~\ref{fig:twofreq} (right side). The PCA reconstruction of 
the data implements the lowest 14 eigenfeatures in the PCA decomposition 
of the entire Stokes-$Q$ map. This order of truncation does a good job
at removing the polarized fringes, however it also appears to clip
some of the finer spectral and spatial information in the \ion{He}{1} 
triplet from the reconstructed data. This is noticeable, for example,
in the appearance of the dark absorption feature located approximately 
at $Y=155$, in the red component of \ion{He}{1} 1083\,nm (spanning
approximately between $X=120$ and $X=140$). The reconstructed data shown
on the right side of the figure is a modification of the PCA expansion 
used for the left side example, which was based on the characteristics
of the eigenfeatures described above. The subset $\{1,4,14\}$ was
removed from the truncated PCA expansion, while we added the subset
$\{16,18\}$. Such operation indeed helped reducing the H$_2$O spectral 
signature from the scan. It was also able to capture more of the
spectral and spatial complexity of the \ion{He}{1} triplet, while at 
the same time preserving the quality of the removal of the polarized 
fringes.



\section{Conclusions}

The problem of the identication and removal of instrumental artifacts,
such as polarized fringes, from spectro-polarimetric maps lends itself
naturally to a treatment by pattern recognition methods, e.g., 
Principal Component Analysis (PCA). In this paper, we have presented the 
2D PCA algorithm of \cite{Ya04}, which appears to outperform other 
PCA strategies in the tests of the authors of that work. 

However, it is interesting to comment briefly also on the more 
traditional implementation of PCA to face recognition \citep{TP91}. 
In this alternate approach, each $m\times n$ image matrix $\tens{A}_i$ 
is rearranged into a $(m\times n)$-vector $\bm{B}_i$. The dataset of 
$N$ images can then be represented by the $(m\times n)\times N$ matrix 
$\tens{B}=\{\bm{B}_i\}_{i=1,\ldots,N}$, and the PCA covariance matrix is
then built as
\begin{equation} \label{eq:alternate}
\tens{C}=\frac{1}{N} \sum_{i=1}^N
	(\bm{B}_i-\bar{\bm{B}})^T (\bm{B}_i-\bar{\bm{B}})\;,
\qquad \bar{\bm{B}}=\frac{1}{N} \sum_{i=1}^N \bm{B}_i\;.
\end{equation}
We note that $\tens{C}$ so defined is an $N\times N$ matrix, and that 
both the spatial and spectral dimensions of the original images 
$\tens{A}_i$ have been contracted in order to compute it. This approach 
has the
notable advantage of producing eigenfeatures that resemble the original
images \cite[see][for details]{TP91}, which provide a direct visual
aid for the selection of the orders that isolate the spectral data from
the fringe background. On the other hand, the high level of data
compression tends to produce a much slower convergence of the PCA
reconstruction series than for the 2D PCA algorithm. The alternative of 
using the products 
$(\bm{B}_i-\bar{\bm{B}}) (\bm{B}_i-\bar{\bm{B}})^T$
for the definition of $\tens{C}$ (see also Sect.~\ref{sec:PCA}) 
would preserve both spatial and spectral information of the covariance 
of the original dataset. On the other hand, this would often create 
a matrix that is too big to be diagonalized efficiently for the typical 
values of the image dimensions, $m$ and $n$. That is why the 
2D PCA algorithm appears more suitable for our problem, despite the fact 
that the order selection may be more cumbersome with this approach. We 
defer a more detailed study of the potential and limitations of the 
traditional PCA approach to future work.

The strategy to make PCA succeed in the removal of polarized fringes
from spectro-polarimetric data is ideally to guarantee the presence in 
a dataset of widely diverse realizations of a line's polarization profiles 
over a practically time-independent fringe background. In fact, this 
determines the condition of non-correlation between the spectral signal 
and the polarized fringes, which is at the very basis of the working 
concept of PCA \citep{Jo02}.

For the \ion{He}{1} 1083\,nm data presented in this work, this condition 
is often met, and as expected PCA manages to separate rather well the 
spectral line information from the fringe background. The \ion{Si}{1} 
line is always prominent in the solar intensity spectrum. However, this 
should not always be the case for polarization, e.g., for quiet-Sun 
observations. So one can hope that the same strategy will work with 
that line as well, at the condition that a sufficient diversity of 
polarization signals over the fringe background can be acquired during 
an observation. In the active-region data that we have analyzed,
instead, the polarization signals of the \ion{Si}{1} line are always
typically at the level of 10-20\% for both linear and circular 
polarization, and thus the subtraction of the fringe background in that
spectral range fails consistently. This problem is greatly mitigated
by the fact that the typical amplitude of the fringes is relatively 
small ($\sim 0.2\%$) in our data, compared to the observed amplitudes 
of the \ion{Si}{1} polarization, and so the polarization profiles of the
line are not significantly affected by the fringe background.


The results presented in this paper point in the direction of revising 
the way that the acquisition of science and calibration data should be 
planned for a typical spectro-polarimetric observing run. Looking at 
different targets on the quiet Sun, while maintaining the same 
configuration of the spectrograph in order to preserve the stationarity 
of the fringe background, could turn out to be a fundamental addition 
to all observing programs. Lamp flat fields may also be an important 
tool for ``fringe calibration'', as they could be used to augment the 
set of fringe data de-voided of spectral line signals for the purpose 
of ``training'' the PCA in the identification 
of fringes. Of course, these lamp flats should ideally be taken under 
the same optical configuration of the spectrograph as used for a 
given observation, which may not always be possible.


\acknowledgments
The authors enjoyed several stimulating discussions with 
A.~L\'opez Ariste (TH\'eMIS-CNRS, France) and A.~Asensio Ramos 
(IAC, Spain) on the problem of removal of polarized fringes by PCA
and other methods. They are also grateful to HAO colleagues L.~Kleint 
and R. Centeno Elliot, D.~Elmore (NSO), and J.~Stenflo 
(ETH, Switzerland), for helpful comments.

\end{document}